\documentclass[aps,prd,twocolumn,superscriptaddress,nofootinbib,showpacs]{revtex4-1}

\usepackage[pdftex]{graphicx}
\usepackage{epsfig}
\usepackage{amsmath}
\usepackage{amssymb}
\usepackage{dcolumn}
\usepackage{multirow}

\newcommand{\beq}{\begin{equation}}
\newcommand{\eeq}{\end{equation}}
\newcommand{\bear}{\begin{eqnarray}}
\newcommand{\eear}{\end{eqnarray}} \newcommand{\ba}{\begin{array}}
\newcommand{\ea}{\end{array}}
\newcommand{\lae}{\begin{array}{c}\,\sim\vspace{-1.7em}\\< 
\end{array}}

\begin{document}

\title{Indirect Probes of Supersymmetry Breaking in the JEM-EUSO Observatory} 
\author{Ivone F.~M. Albuquerque} 
\email{ifreire@if.usp.br}
\author{Jairo Cavalcante de Souza} 
\email{jairocavalcante@gmail.com}
\affiliation{Instituto de F\'{i}sica, Universidade de S\~ao Paulo, S\~ao Paulo, Brazil}

\date{\today}

\begin{abstract}
In this paper we propose indirect probes of the scale of supersymmetry breaking,  
through observations in the Extreme Universe Space Observatory onboard Japanese Experiment
Module (JEM-EUSO). 
We consider scenarios where the lightest supersymmetric particle
is the gravitino, and the next to lightest (NLSP) is a long lived slepton. We demonstrate 
that JEM-EUSO will be able to probe models where the NLSP decays,
therefore probing supersymmetric breaking scales below $5 \times 10^6$~GeV. The observatory field of view will be large enough to detect a few tens of events per year,
depending on its energy threshold.
This is complementary to a previous proposal~\cite{abc} where it was shown
that 1~Km$^3$ neutrino telescopes can directly probe this scale.
NLSPs will be produced by the interaction of high energy neutrinos in the Earth.
Here we investigate scenarios where they subsequently decay, either in the atmosphere after escaping the Earth
or right before leaving the Earth, producing taus. These can be detected by JEM-EUSO and have
two distinctive signatures: one, they are produced  in the Earth and go upwards in
the atmosphere, which allows discrimination from atmospheric taus and, second, as NLSPs are 
always produced in pairs, coincident taus will be a strong signature for these events.
Assuming that the neutrino flux is equivalent to the
Waxman-Bahcall limit, we determine the rate of taus from NLSP decays reaching JEM-EUSO's
field of view. 
\end{abstract}


\maketitle

\section{\label{intro} Introduction}

Probes of the origin and stability of the weak scale and proposed solutions to the
standard model hierarchy problem is now underway at the Large Hadron Collider
(LHC). On the theoretical side, the supersymmetric theory (Susy) arises as
a solution with no significant deviation from the standard model (SM) in relation to
electroweak precision observables. 
It has been shown~\cite{abc,abcd} that neutrino telescopes can
probe the scale of supersymmetry breaking  ($\sqrt{F}$) or
of Universal Extra Dimensions scenarios~\cite{abckk}. While 
R parity symmetry ensures that the lightest 
supersymmetric particle (LSP) is stable, the identity of the LSP is determined by 
$\sqrt{F}$. If Susy is broken at $\sqrt{F}>10^{10}$~GeV the LSP is typically 
the neutralino, if $\sqrt{F}<10^{10}$~GeV it is typically the gravitino.

There are many Susy scenarios where $\sqrt{F}$ is low, among which Gauge Mediation
Susy Breaking models~\cite{gmsb}. In these scenarios the NLSP is a charged slepton, 
typically a right-handed stau, and it's lifetime is given by~\cite{abc}:

\beq
 c\tau = \left(\frac{\sqrt{F}}{10^7{\rm~GeV}}\right)^4\, 
\left(\frac{100~{\rm GeV}}{m_{\tilde{\tau}}}
\right)^5\,10~{\rm km}~,
\label{eq:ctau}
\eeq
where ${m_{\tilde\tau}}$ is the stau mass. It was shown \cite{abc,abcd} that 
Km$^3$ neutrino telescopes can directly probe the Susy breaking scale, when one
consider scenarios where $5 \times 10^{6} < \sqrt{F} < 10^{8}$~GeV. In these
scenarios, NLSPs produced in very high energy collisions will travel very long distances
before decaying. A detectable flux of NLSPs can be produced
by the interaction of the diffuse flux of high energy neutrinos with the Earth.

Here we consider \cite{tese} a complementary Susy breaking scale region, with
$\sqrt{F} \lae 10^{7}$~GeV, which implies that the NLSP will decay after a short travel. After being 
produced by neutrino interactions in the Earth, a good
fraction of these particles will decay inside the Earth or in the atmosphere, after escaping the Earth.
In this work we consider NLSPs decay in the atmosphere (or right before reaching it). In a complementary
investigation~\cite{tese,decice} we analyze the $\sqrt{F}$ region where they decay inside the Earth
and might be detected by multi-km$^3$ neutrino telescopes.
Once the NLSPs decay, taus ($\tau$s) will be produced and can be detected by fluorescence telescopes. We show 
that the JEM-EUSO observatory \cite{euso} can probe these scenarios,
where NLSP decays will yield a few tens of events per year, depending
on its  lower energy threshold.

We do consider the regeneration process, where $\tau$s decay into $\tau$ neutrinos which in 
turn will charge current interact inside the Earth producing new
$\tau$s. However these suffer large energy degradation, and only a small
fraction of these events reach the detector with significant energy. 

As in \cite{abc}, the crucial idea in probing $\sqrt{F}$ is that although the NLSP 
production cross section is much smaller than the one for SM lepton
production, the NLSP range in the Earth is 
much larger than for a standard lepton. As shown in 
\cite{abcd}, the NLSP energy loss is much smaller than the one for muons or $\tau$s. 
Here, when NLSP decays are considered, the NLSPs can be produced even
further away from the detector, which implies that an extra decaying volume is gained. 
Also, as the NLSPs are always produced in pairs, a considerable
fraction of the observable events will consist of a pair of  $\tau$s in the
atmosphere. These coincident $\tau$s will be a distinctive indirect signature
of NLSPs. Another strong signature comes from the fact that these $\tau$s
emerge from the Earth and go upwards in the atmosphere, which
discriminates them from down going atmospheric $\tau$s.

In this paper we describe our analysis, where we developed a Monte Carlo simulation to
investigate $\tau$ events generated from NLSP decays. The first steps,
reviewed in the next section, reproduce the analysis done in~\cite{abcd}, where NLSP production,
propagation and energy loss are described in details. 
Subsequently, in section \ref{sec:rate}, we describe our simulation of NLSP decays and produced $\tau$
propagation. The signatures and rates of these events in the
JEM-EUSO observatory are determined in section~\ref{sec:euso}, as well as
their discrimination from the background. Finally we discuss our
results and state conclusions.

\section{\label{sec:nlsp} NLSP PRODUCTION}

Here we consider NLSP production by a diffuse flux of high energy neutrinos 
interacting in the Earth. The neutrino flux as well as
the NLSP production cross section and propagation are determined as described in
detail in \cite{abcd}, and are used as the first
steps in our Monte Carlo simulation. In the next section we describe the original
part of our work, where NLSP decays are included, and the produced $\tau$ 
rates in the JEM-EUSO observatory are determined.

Although the diffuse flux of high energy neutrinos that reaches the Earth is yet unknown, 
there are several estimates
of its upper limit. Waxman and Bahcall (WB) \cite{wb} determined such a limit
based on the observed cosmic ray flux, since neutrinos are produced
from pion decays, which in its turn are produced from proton interactions.
Considering optically ``thin'' sources, which would allow most 
of the protons to escape, they determine the neutrino spectrum as 
\beq
\left(\frac{d\phi_\nu}{dE}\right)_{\rm WB} = \frac{(1-4) \times
10^{-8}}{E^2} {\rm GeV~ cm^{-2} s^{-1} ~sr^{-1}}~,
\label{wblimit}
\eeq 
where the allowed interval depends on the cosmological evolution of the sources.
Here we adopt the upper value of the WB limit as the cosmological neutrino flux
that reaches the Earth. Our results can be translated to other neutrino fluxes by properly rescaling the 
WB limit. The initial neutrino flux contains both muon and electron neutrinos, in a $2:1$
ratio. However we note that the initial neutrino flavor does not alter our results,
since its interaction will always produce a left-handed slepton that will always
immediately decay, having the right-handed slepton as a final product. For the
same reason a large mixing of the cosmogenic neutrino flux does not modify our
results.

Once the neutrino flux is defined, the NLSP production cross section should be 
determined. We follow the cross section calculation described in detail in \cite{abcd}
and reproduce their results. In short the $\nu N \to \tilde{l}_L \tilde{q}$
process is analogous to the SM charged current interaction with $\tilde{q}$ being an
up or down type squark and, in the t-channel, the mediator is the chargino. The 
sub-dominant
process with a neutralino exchange is also included in the cross section calculation.
As a result of this interaction a $\tilde{l}_L$ and a $\tilde{q}$ will be produced.
These will immediately decay in a chain that will always end with the production of
two NLSPs, typically the right-handed stau ($\tilde{\tau}$). We give our results considering the mass
of the chargino and of the left handed slepton respectively as $m_{\tilde w}=250$~GeV
and $m_{\tilde{l}_L}=250$~GeV, of the NLSP as $m_{\tilde{\tau}}=150$~GeV and three
possibilities for the squark mass, $m_{\tilde q}=300$, ~$600$ or
$900$~GeV. Note that the LHC constrains $m_{\tilde q}$ to larger
values when considering specific scenarios that have the neutralino as the LSP
\cite{pdg}, which is not our case. There are constrains on 
$m_{\tilde{\tau}}$  from
big-bang nucleosynthesis~\cite{stbbn}.

The probability of a neutrino interacting in the Earth as well as the NLSP propagation
through this medium, depend on the Earth density profile model. We use the model \cite{earth}
described in detail in \cite{gandhi}. The lepton production cross section,
from neutrino nucleon interaction is also determined in \cite{gandhi}. 

Once the NLSP is produced it will propagate through the Earth and lose energy. The main
process for NLSP energy degradation is photo-nuclear interactions. However, as shown in 
\cite{abcd,ina}, all radiative losses are suppressed due to the NLSP heavier mass
when compared to standard leptons.

NLSPs will always be produced in pairs and therefore, if one assumes the $\sqrt{F}$ 
region where they do not decay, they will have a very distinctive signature in neutrino
telescopes. As a cross check of our simulation, we reproduced the NLSP rate and energy
distribution in Km$^3$ neutrino telescopes as determined in \cite{abc,abcd}.

\section{\label{sec:rate} NLSP Decay} 

\begin{figure}
\includegraphics[width=\linewidth]{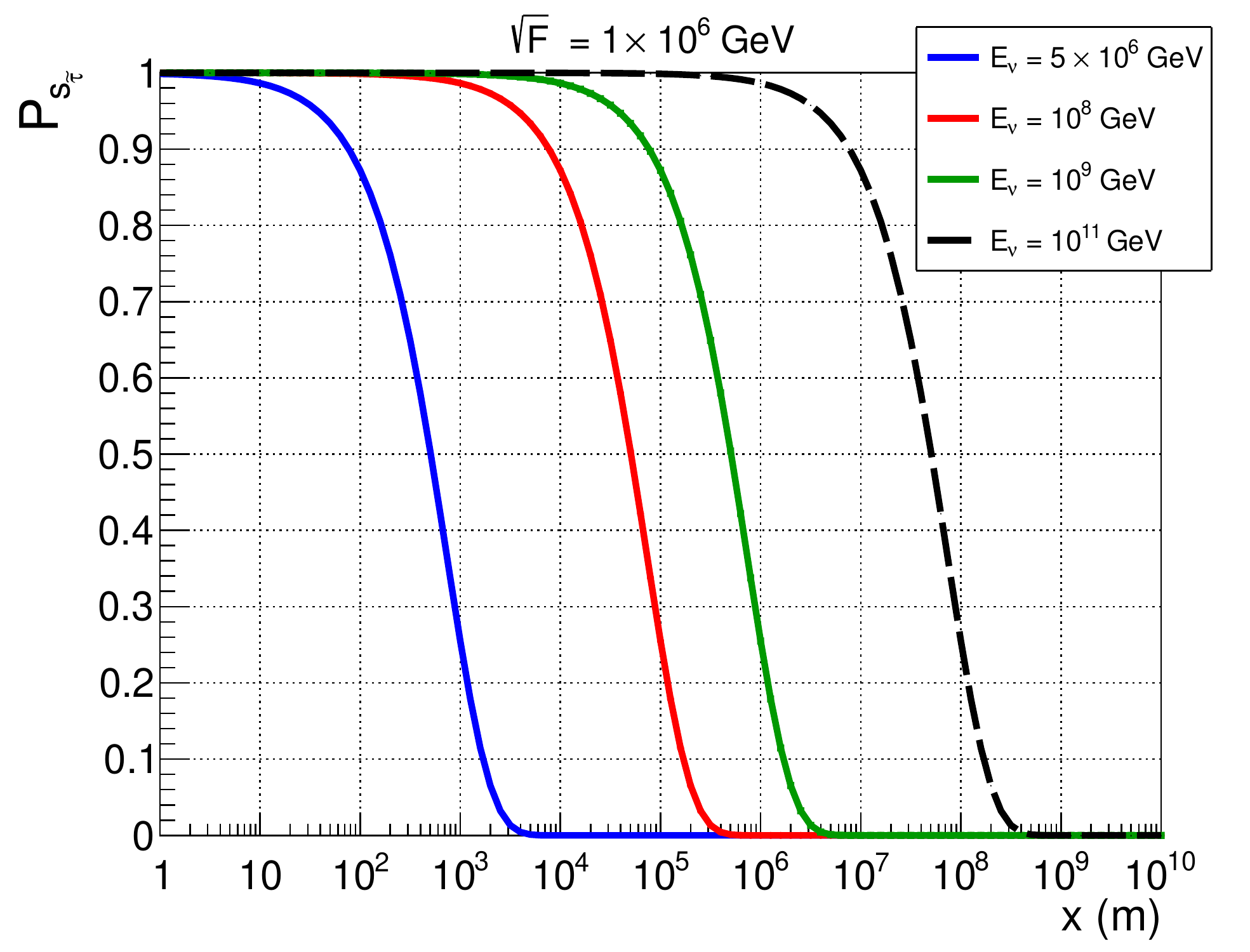}
\caption{NLSP survival probability as a function of traveled distance, for different
neutrino energies (as labeled) and $\sqrt{F}=10^6$~GeV.}
\label{fig:sdec}
\end{figure}

\begin{figure}
\includegraphics[width=\linewidth]{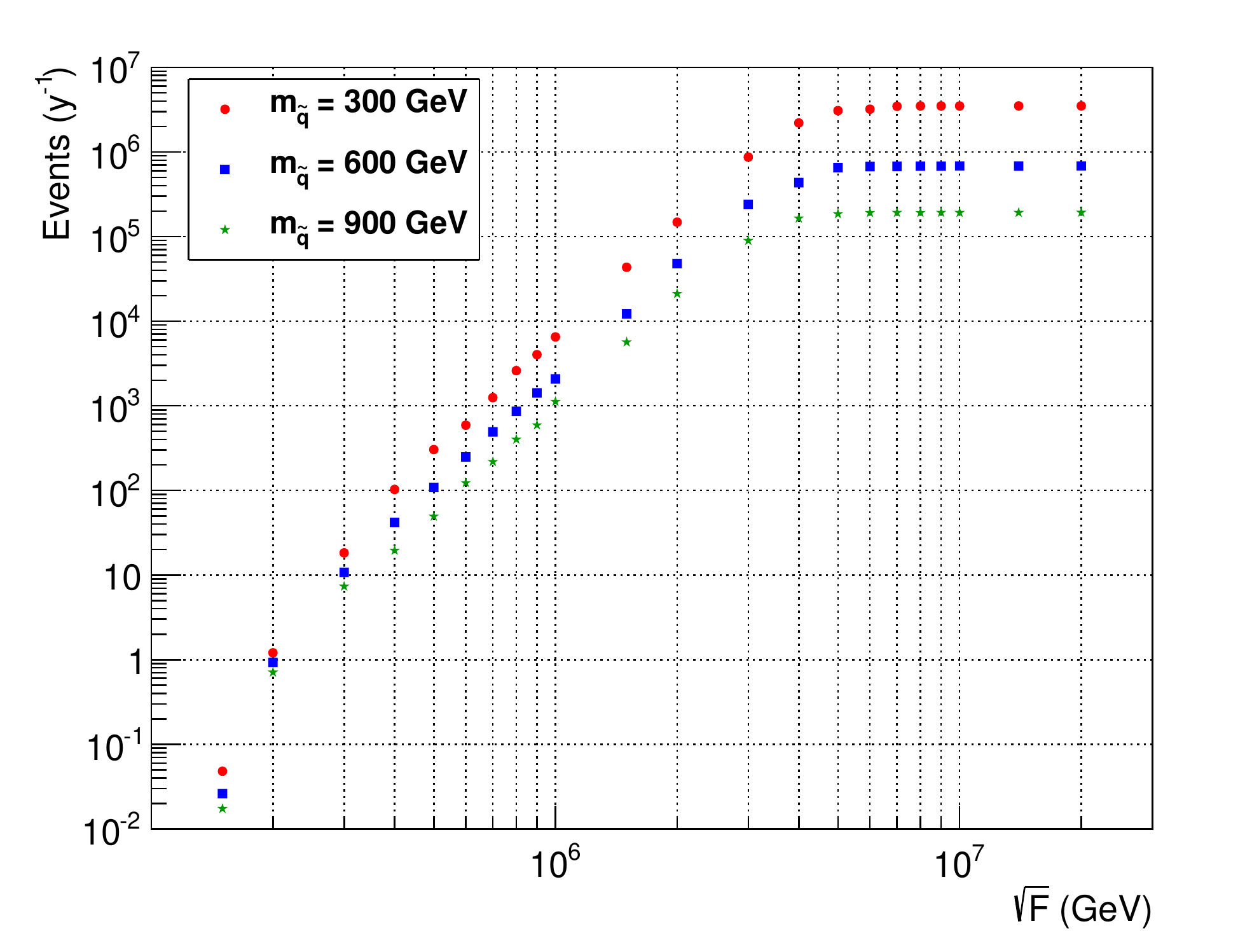}
\caption{Fraction of NLSP events in JEM-EUSO's field of view (FOV) versus
  $\sqrt{F}$ for $m_{\tilde q} $ reference values. NLSPs decay for
$\sqrt{F} \lae 5 \times 10^6$˜GeV. }
\label{fig:sqf}
\end{figure}

The NLSP survival probability 
\beq
P_s(x) = \exp(-m_{\tilde{\tau}} x/E_{\tilde{\tau}} c\tau)
\label{eq:decp}
\eeq
is shown in Figure~\ref{fig:sdec} as a function of the traveled distance $x$,
for different neutrino energies $E_{\nu}$ and $\sqrt{F}=10^6$~GeV, where
the initial NLSP energy  is typically $E_{\nu}/6$~\cite{abcd}.
The NLSP  production energy threshold is $\sim 10^6$~GeV, and defined by the 
$\tilde{q}$ and the left-handed $\tilde{l}_L$. At these energies, the distance
$\gamma c \tau$ is determined by Equation~\ref{eq:ctau} and NLSPs decay
when $\sqrt{F} \lae 5 \times 10^{6}$~GeV. This feature is shown in 
Figure~\ref{fig:sqf}. The decay channel is $\tilde{\tau} \to \tau + \tilde{G}$, 
where $\tilde{G}$ is the gravitino and its mass is much smaller than the $\tau$ 
mass. 

We developed a Monte Carlo simulation, where $\vartheta(10^5)$ events where generated, corresponding to neutrinos reaching the Earth, 
isotropically distributed both in energy and impinging
direction. These events are normalized by the WB limit. 
The NLSP production, propagation and energy loss were simulated,
reproducing the analysis shown in ~\cite{abcd} (see Section~\ref{sec:nlsp}). 

The NLSP decay point is chosen from the decay probability distribution
(essentially $1 - P_s$), 
and the generated $\tau$ center of mass isotropic angular distribution is boosted into the 
laboratory frame. To a good approximation, it will follow the same
direction as its parent NLSP.  As two NLSPs are always generated in
pairs, two $\tau$s will be produced. A fraction of these $\tau$s subsequently decay,
always generating a $\nu_\tau$, which can charge current interact producing
a new $\tau$. 
We determine the fraction of the $\tau$ energy carried by the $\nu_\tau$ as in 
\cite{crotty,dutta}. This regeneration process can happen a few times depending 
on the $\tau$ energy degradation.  As mentioned before, this process
will degrade the $\tau$ energy and most of them will not be detectable.

\section{\label{sec:euso} Signatures and Event Rates in the JEM-EUSO Observatory} 

In order to determine the feasibility of NLSP indirect detection, where $\tau$s originated
in NLSP decays would be probed by a large fluorescence telescope, we
have to consider basic detector features. The JEM-EUSO telescope \cite{euso}
will observe fluorescence light emitted from an extensive air shower
created by the interaction of a high energy particle in
the atmosphere. It will orbit the Earth at an altitude of about
430~Km, yielding a detection area of $\sim 2 \times 10^5$~Km$^2$ with
a 250~Km circular radius at the Earth surface. It is scheduled to be launched in 2016.

At these energies, NLSP produced $\tau$s can be observed by JEM-EUSO
mainly through the
shower created once they decay. NLSP direct detection is hard, since
the energy of its emitted fluorescence light is much less than the detector threshold,
which is projected to be around $10^{19}$~eV.

\begin{figure}
\includegraphics[width=\linewidth]{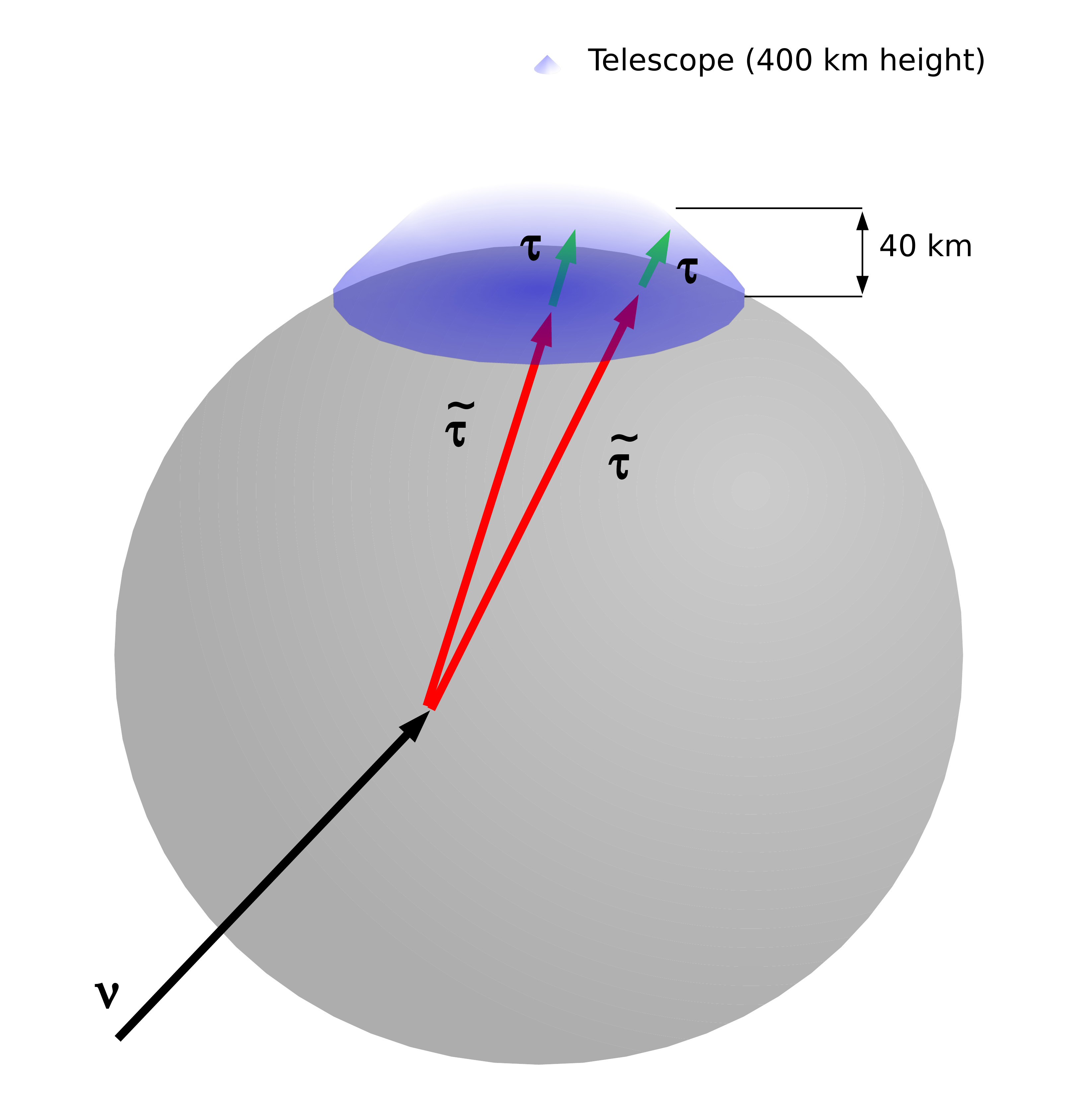}
\caption{Representation of JEM-EUSO detection volume as a frustum of
  a cone, with a 40~Km height corresponding to the atmosphere
  altitude. NLSPs originated from neutrino interactions will decay in
  the atmosphere, producing observable $\tau$s. }
\label{fig:fov}
\vspace*{-.5cm}
\end{figure}

In order to determine the rate of observable events, we approximate
the JEM-EUSO detection volume as a frustum of a cone, which represents
the detector field of view. This is shown in figure~\ref{fig:fov}. The frustum's height corresponds to the
atmosphere altitude which is taken as 40~Km. 

As described in the previous section, we simulate NLSP production from neutrino
interactions in the Earth, their propagation and energy loss and
finally their decay. The neutrino energy ranges
from $\sim 10^6$ to $\sim 10^{12}$~GeV, where the lower limit corresponds to the NLSP production
threshold. As the NLSPs are always produced in pairs, two $\tau$s will
be produced from their decay. Although the pair of NLSPs
travel in parallel, they can decay at different times, and each $\tau$
can be produced at different positions. Each production point is
independently chosen from the decay probability distribution.

The $\tau$ energy loss considers both ionization and
radiation processes, where the latter includes loss due to
bremsstrahlung, pair production and photo-nuclear interactions
\cite{pdg,abcd}.   We also follow the NLSPs which do
not decay inside the Earth and reach the atmosphere, computing their
energy and flux. The NLSP energy loss to the atmosphere is negligible.

\begin{figure}
\includegraphics[width=\linewidth]{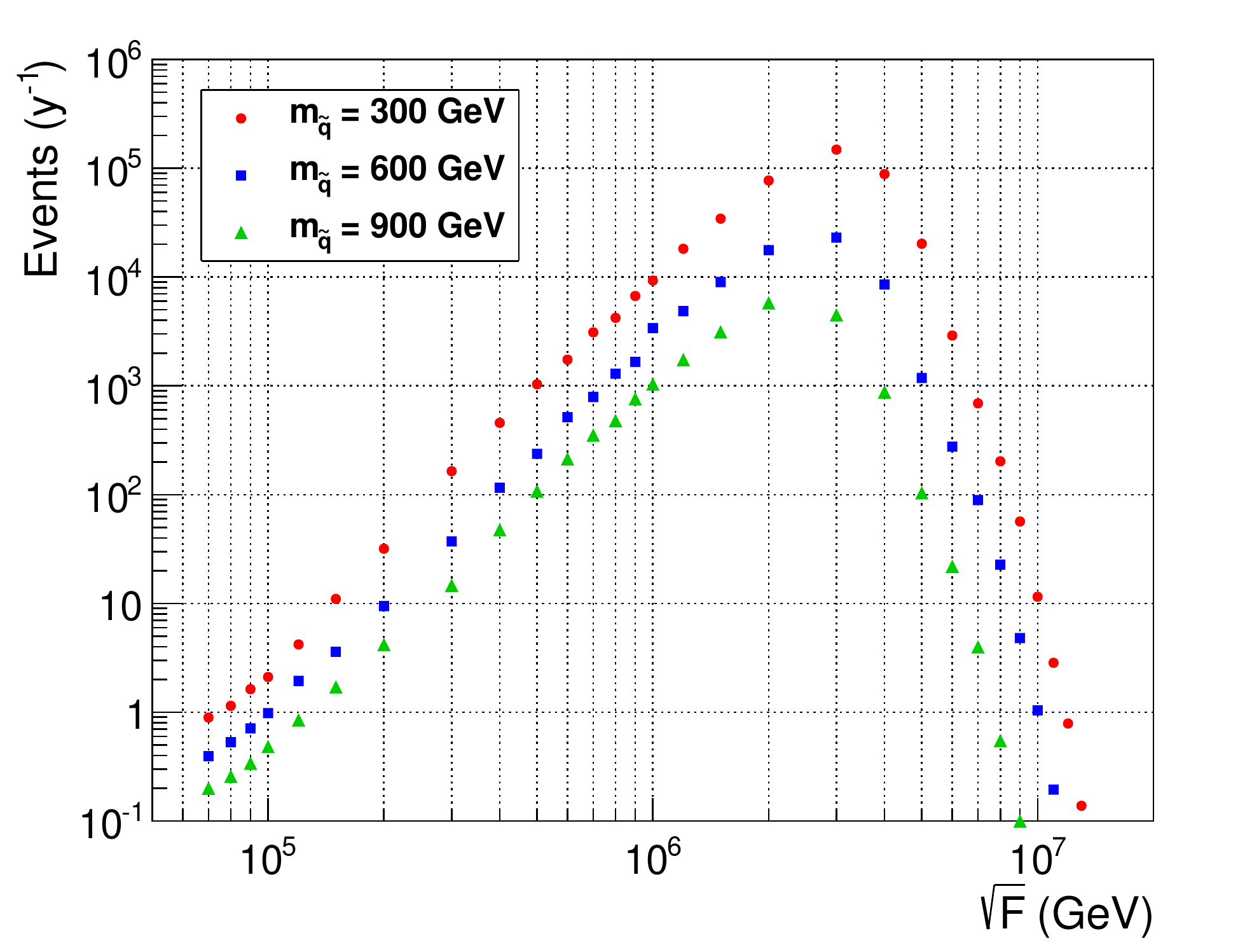}
\caption{NLSP coincident decays in JEM-EUSO FOV as a function of $\sqrt{F}$.}
\label{fig:sqrtf}
\end{figure}

In summary we compute the $\tau$ production point and initial energy
distribution, as well as its decay point and energy distribution. These yield the number of $\tau$s which decay
in or propagate through the detector field of view, and corresponding
energy distributions. We also compute the number of coincident $\tau$ decays
in the field of view. All results are normalized by the WB neutrino
flux.

\begin{figure}
\includegraphics[width=\linewidth]{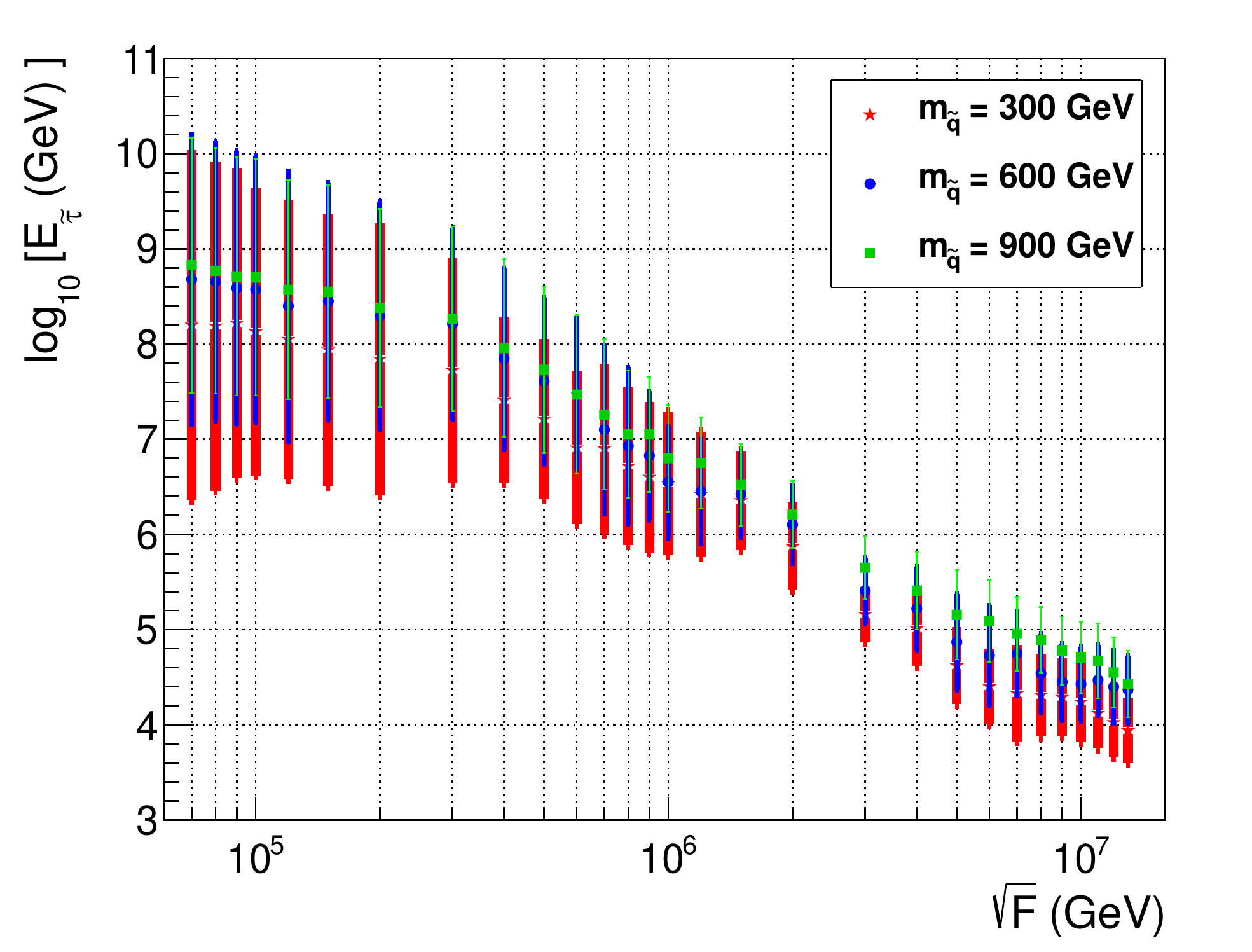}
\caption{Average energy of events shown in
Figure~\ref{fig:sqrtf}. Error bars represent standard deviation of each
energy distribution.}
\label{fig:en}
\end{figure}

 \begin{figure}
\includegraphics[width=\linewidth]{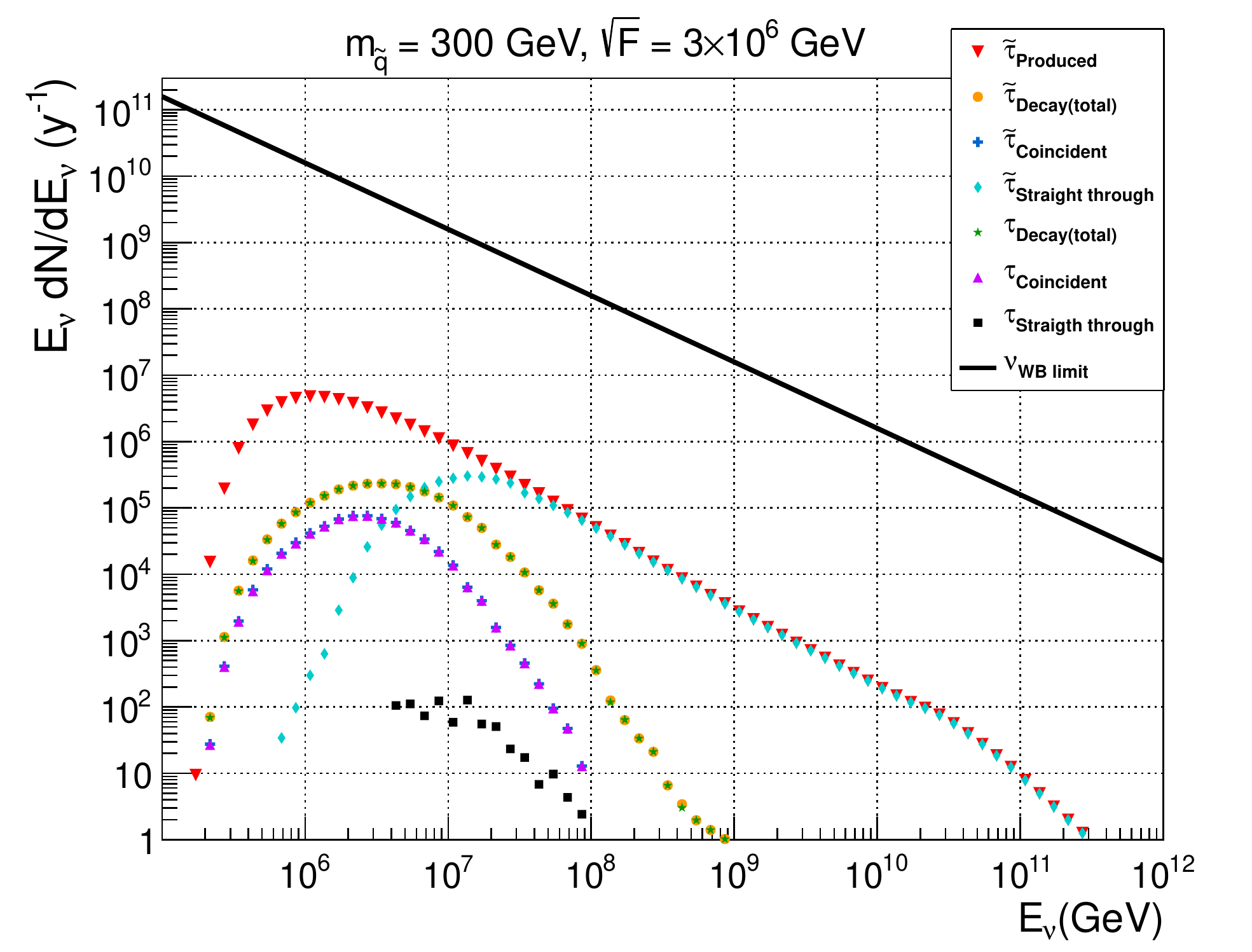}
\caption{Events per year in JEM-EUSO FOV. Here $m_{\tilde q} =
  300$~GeV and $\sqrt{F} = 3 \times 10^6$~GeV, such that the number of
events is maximized. The circles correspond
to NLSPs (represented as $\tilde\tau$) and triangles to $\tau$s. The NLSP ($\tau$)
that decay are in yellow (green), the ones that go through the FOV without
decaying in light blue (black), the ones which both NLSPs produced as a pair (or both $\tau$s)
decay in the FOV in blue (purple). As a reference the total number
of NLSPs that are produced in the direction of the FOV are shown in
red and the WB limit as a black line. Other values of $m_{\tilde q}$
will have similar curves but with less events.}
\label{fig:evfov} 
\end{figure}
 
Our results are presented as a function of
 $\sqrt{F}$. Figure~\ref{fig:sqrtf} shows the number of coincident
 NLSP decays per year in JEM-EUSO field of view (FOV), while Figure~\ref{fig:en} shows
 the NLSP average decay energy. As can be seen, the number of events is
 maximized for $\sqrt{F} = 3 \times 10^6$~GeV for both 300 and 600~GeV
squark masses and $2 \times 10^6$~GeV for $m_{\tilde q} = 900$~GeV,
due to its higher production energy threshold. However at these values
of $\sqrt{F}$, the NLSP average decay energy is lower than the
detection threshold, and increases for lower
$\sqrt{F}$ values. We will further discuss this issue.

Figure~\ref{fig:evfov} shows the number  of events per year in JEM-EUSO
FOV as a function of the parent neutrino energy, for $\sqrt{F} = 5
\times 10^6$~GeV which maximizes  the number of
events. It shows the NLSPs
($\tilde\tau$) and $\tau$s that decay, the ones that go through the FOV without
decaying, the ones which both NLSPs produced as a pair or both $\tau$s
decay coincidently in the FOV. As a reference, the figure also shows the total number
of NLSPs that are produced in the direction of the FOV and the WB
limit. This figure shows results assuming $m_{\tilde q} = 300$~GeV,
where for the other $m_{\tilde q}$ values the shape of the curves are
very similar but contain less events.  The total rate of events
in JEM-EUSO  per year, for our nominal value of maximum atmosphere
height of 40 Km, for the three values of $m_{\tilde q}$, are shown in 
Table~\ref{tab:ev1y}.

\begin{figure}
\includegraphics[width=\linewidth,height=20.cm]{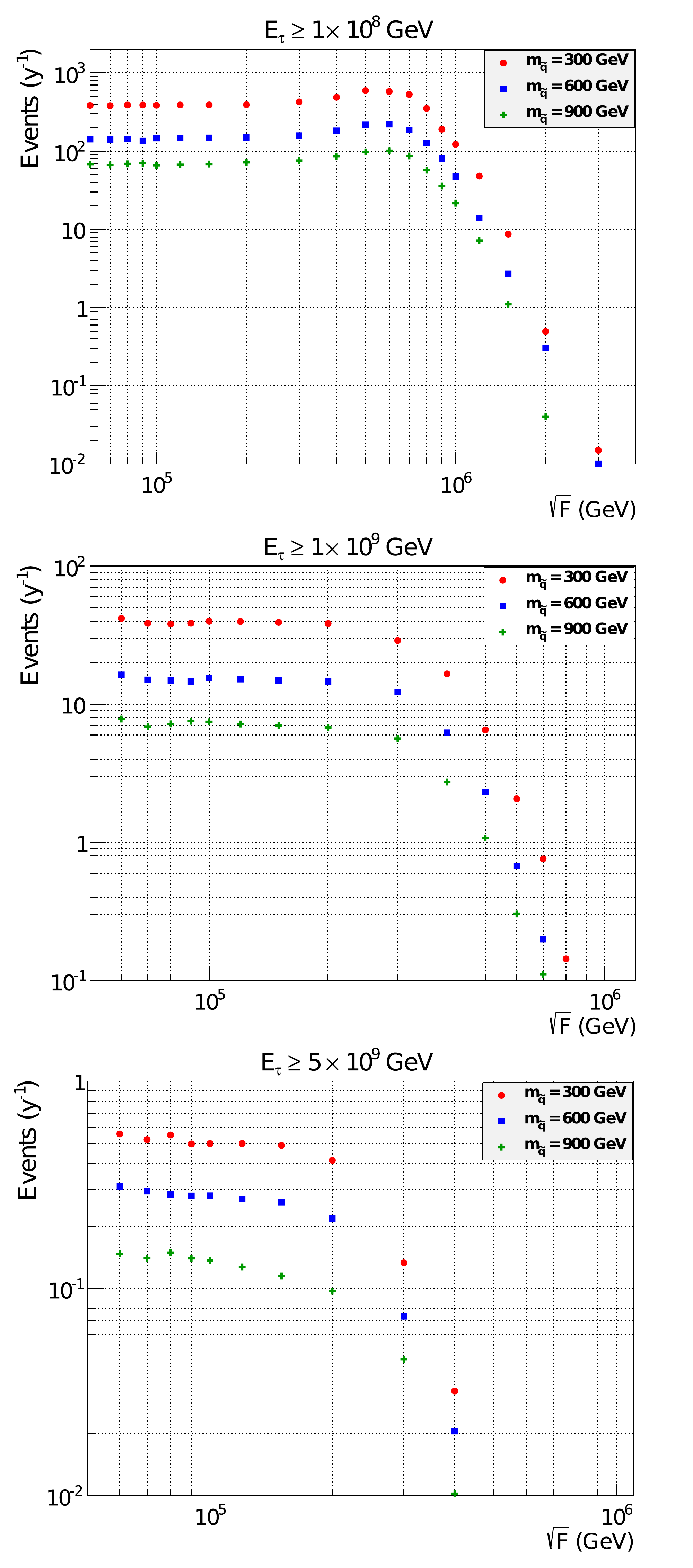}
\caption{Total number of $\tau$ decays in JEM-EUSO FOV, requiring a
  minimal energy of $10^{8}$~GeV (top plot); $10^9$~GeV (middle plot)
  and $5 \times 10^{9}$~GeV (bottom plot).}
\label{fig:encut} 
\end{figure}

\begin{table}
\begin{center}
\vspace{-8pt}
\caption{{\small Number of NLSP events per year in JEM-EUSO FOV, where
 $\sqrt{F} = 3 (2) \times 10^6$~GeV for  $m_{\tilde q} = 300, 600
 (900)$~GeV. For $m_{\tilde q} = 300$~GeV these events correspond to
 the ones shown in Figure~\ref{fig:evfov}. ``Total'' is the
 total number of NLSPs produced in the direction of the FOV; ``Straight Through''
refers to the NLSPs that go through the FOV; ``Decay '' to
 those which decay in the FOV; and ``Pair Decay'' to the ones where
 the NLSP pair decays in coincidence in the atmosphere.''}}
\vspace{8.pt}
\begin{tabular}{c|c|c|c|c}
\hline\noalign{\smallskip}
\vspace{0.25pt}\\
$m_{\tilde{q}}$ & Total              & Straight Through         &   Decay  & Pair decay \\
\hline\noalign{\smallskip}
 300            &  $1\times10^{7}$   &  $8.7\times10^{5}$ &   $5.2\times10^{5}$  &  $1.5\times10^{5}$ \\
\hline\noalign{\smallskip}
 600            &  $1.3\times10^{6}$ &  $2.4\times10^{5}$ &   $9.6\times10^{4}$  &  $2.2\times10^{4}$ \\
\hline\noalign{\smallskip}
 900            &  $2.8\times10^{5}$ &  $2.0\times10^{4}$ &   $1.9\times10^{4}$  &  $5.8\times10^{3}$ \\
\hline\noalign{\smallskip}
\end{tabular}
\label{tab:ev1y}
\end{center}
\end{table}

As seen in this table a huge amount of NLSPs should either go through or decay in JEM-EUSO
FOV. As the NLSPs are not detectable due to the low amount of
fluorescence emitted while they transverse the atmosphere, the hope of
probing these events relies on the $\tau$s produced in their
decay. However, although more than $10^5$ $\tau$s will tranverse or
decay in JEM-EUSO FOV, most of these events have
energies below the detector threshold ($\sim 10^{19}$~eV) and will
not be observed unless this threshold is lowered.

In order to determine the event rate that can be observed with an energy 
threshold around the currently projected by JEM-EUSO, we selected
events above arbitrary lower energy  values of $10^8, 10^9$ and $5
\times 10^9$ ~GeV. These are shown in Figure~\ref{fig:encut}  and
the integrated number of events, maximized as a function of $\sqrt{F}$,
in Table~\ref{tab:encut}.

\begin{table}
\begin{center}
{\small 
\caption{Number of $\tau$ events per year in JEM-EUSO FOV after requiring a
minimum $\tau$ energy $E_{\tau} $. $\sqrt{F}$ is such that it
maximizes the number of events. The line corresponding
to $\sqrt{F} = 3 \times 10^6$~GeV is shown for reference and has no
energy requirement.}
\vspace{12pt}
\begin{tabular}{|c|c|c|c|}
\hline
\multicolumn{4}{|c|}{$m_{\tilde{q}}$ = 300 GeV} \\
\hline
$\sqrt{F}$ (GeV)& $E_{\tau}$ (GeV) & $\tau_{Decay(total)}$ & $\tau_{Coincident}$\\
\hline
$3.0\times10^{6}$ &              &  $5.5\times10^{5}$    & $1.5\times10^{5}$  \\
\hline
$6.0\times10^{5}$ &     $\geq 10^8$   & $5.8\times10^{2}$     & $1.0\times10^{2}$ \\
\hline
$1.5\times10^{5}$ &     $\geq 10^9$   & $3.9\times10^{1}$     & $4.2\times10^{0}$  \\
\hline
$1.2\times10^{5}$ &    $\geq 5 \times 10^{9}$ & $5.0\times10^{-1}$     & $1.0\times10^{-2}$\\
\hline
\hline
\multicolumn{4}{|c|}{$m_{\tilde{q}}$ = 600 GeV} \\
\hline
$3\times10^{6}$ &             &  $9.8\times10^{4}$    & $2.2\times10^{4}$  \\
\hline
$6\times10^{5}$ &    $\geq 10^8$   & $2.2\times10^{2}$     & $3.8\times10^{1}$ \\
\hline
$1.2\times10^{5}$ &  $\geq 10^9$   & $1.5\times10^{1}$     & $1.6\times10^{0}$  \\
\hline
$2 \times10^{5}$ &   $\geq 5\times 10^{9}$  & $2.0\times10^{-1}$     & $1.0\times10^{-2}$\\
\hline
\hline
\multicolumn{4}{|c|}{$m_{\tilde{q}}$ = 900 GeV} \\
\hline
$2\times10^{6}$ &              &  $2.0\times10^{4}$    & $5.8\times10^{3}$  \\
\hline
$6\times10^{5}$ &     $\geq 10^8$   &  $1.0\times10^{2}$    & $1.8\times10^{1}$ \\
\hline
$1.2\times10^{5}$ &     $\geq 10^9$   &  $7.2\times10^{0}$    & $8.0\times10^{-1}$  \\
\hline
$1.2\times10^{5}$ &    $\geq 5 \times 10^{9}$  &  $1.3\times10^{-1}$   & $6.0\times10^{-3}$\\
\hline
\end{tabular}
\label{tab:encut}
}
\end{center}
\end{table}

As can be seen from Table~\ref{tab:encut} there is still a reasonable
amount of events for a maximized $\sqrt{F}$ around
JEM-EUSO energy threshold. Above $10^{18}$~eV, 39 (15, 7) $\tau$s
can be detected per year, for a 300 (600, 900)~GeV squark mass. From
these 4 (2, 1) $\tau$ pairs decay in coincidence in the atmosphere, which provides an excellent discriminating signature.
For a threshold of $ 5 \times 10^{18}$~eV, a bit below the nominal projected threshold, 
0.5 (0.2, 0.1) events per year can be observed by JEM-EUSO. 

\begin{figure}
\includegraphics[width=\linewidth]{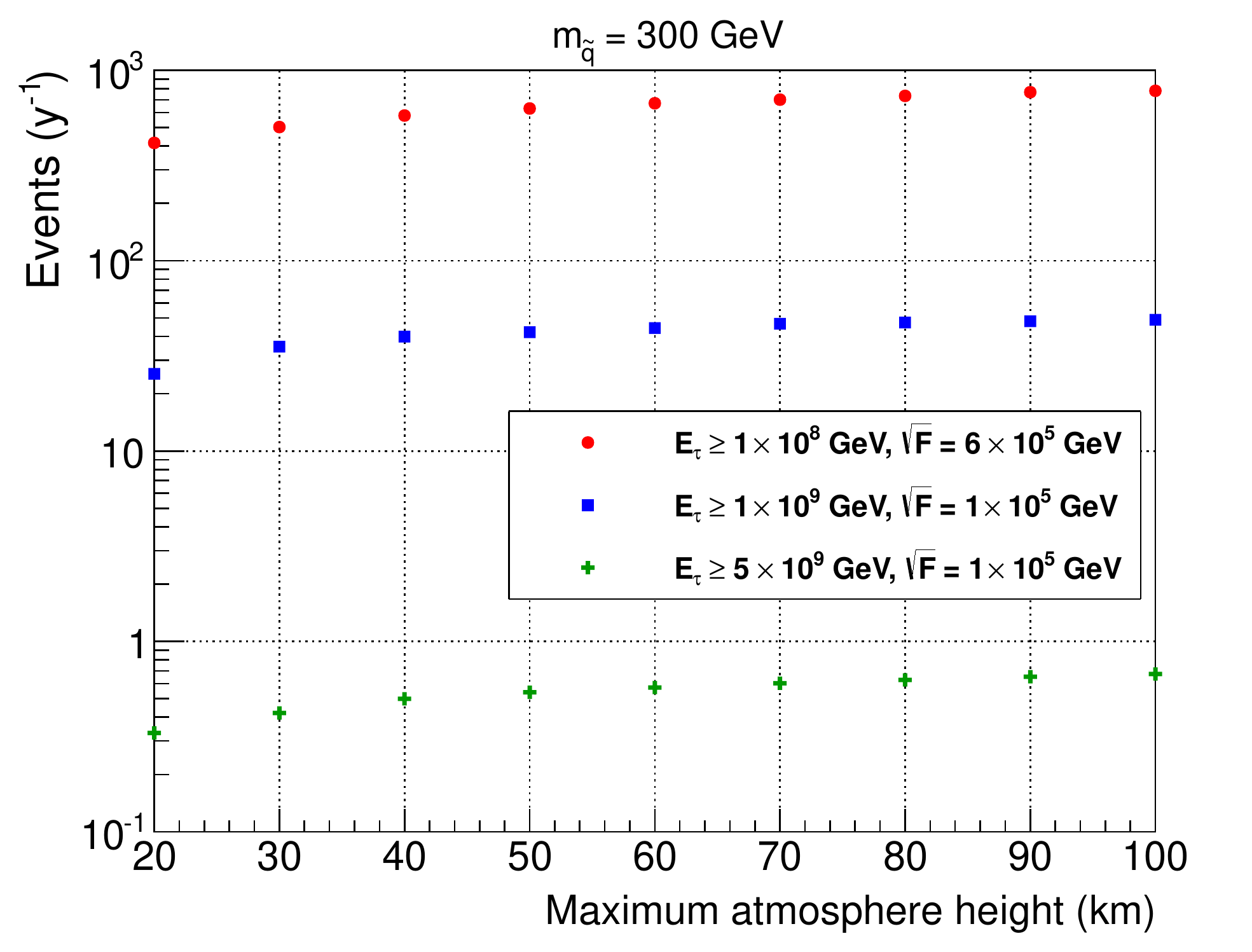}
\caption{Number of $\tau$ decays as a function of
  the maximum atmosphere height, for different values of required minimal
$\tau$ energy.}
\label{fig:evph} 
\end{figure}

In Figure~\ref{fig:evph} we show
the total number of $\tau$ decays as a function
of the maximum atmosphere height, for different values of the required
minimal $\tau$ energy. This allows one to rescale our results
when considering different field of view heights. Although this figure
is shown only for a $m_{\tilde{q}} = 300$~GeV, the rescaling for other
masses is about the same.

\subsection{\label{sec:bckg} Backgrounds}

\begin{figure}
\includegraphics[width=\linewidth]{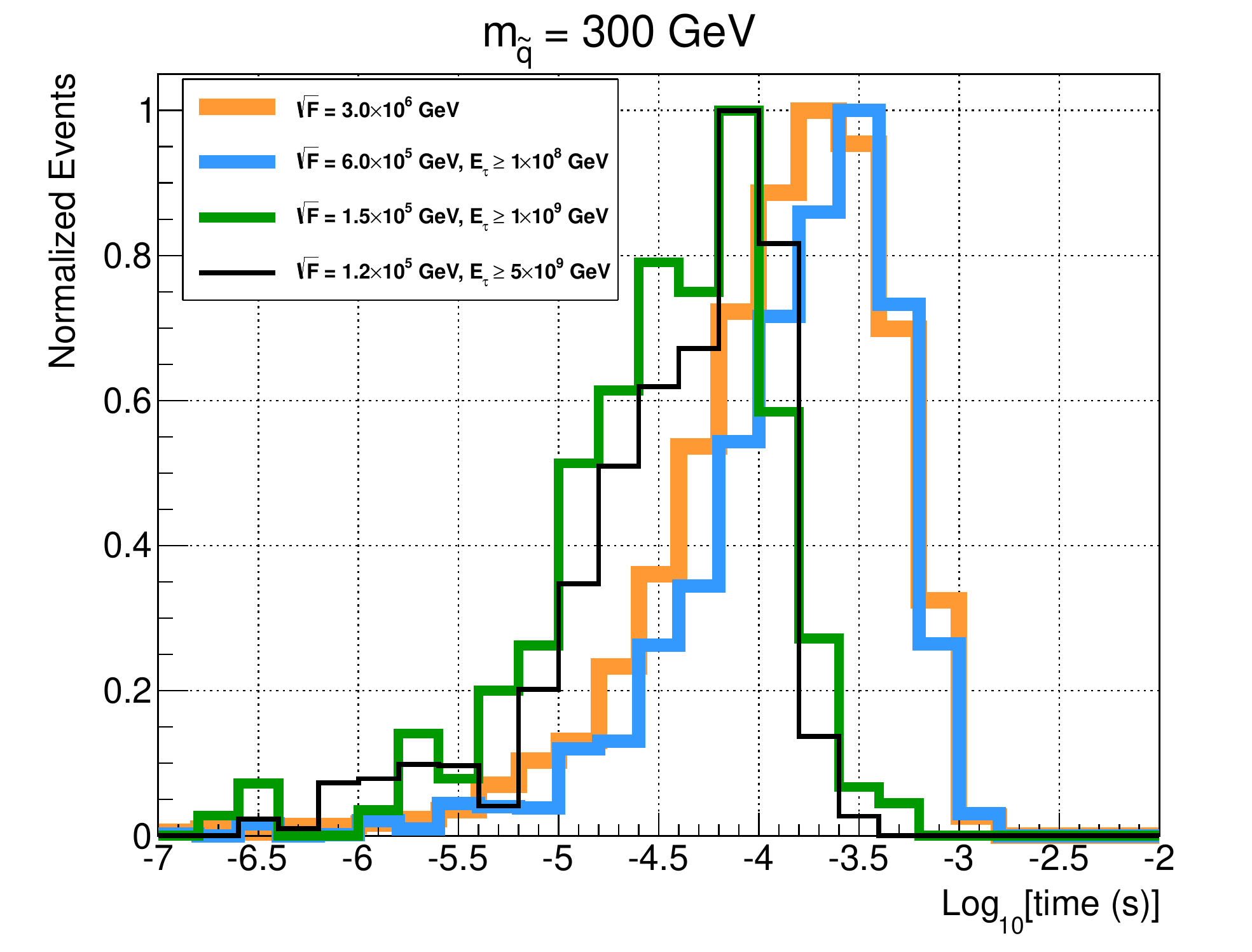}
\caption{Time delay between two $\tau$ coincident decays in the atmosphere. Shown
for arbitrary $\sqrt{F}$ and lower NLSP energy requirements. The
squark mass is 300~GeV but it does not vary considerably for larger values.}
\label{fig:time} 
\end{figure}
 
Under the scenarios we are considering, NLSP decays will have a very
distinctive signature. Since the NLSPs are produced inside the Earth
and decay into $\tau$s that go upwards in the atmosphere, they can 
be discriminated from the more abundant down going cascades produced by normal
ultra high energy cosmic rays.

Also a considerable fraction of the NLSP pairs decay in coincidence in
the atmosphere. Figure~\ref{fig:time} shows the distribution of the
time delay between two
$\tau$ decays in the atmosphere. These are shown for different
$\sqrt{F}$ and energy lower limit requirements. It shows
results for a 300~GeV squark mass, which do not vary significantly
for larger values. The time delay spread as a
function of $\sqrt{F}$ is small, and is distributed around
$10^{-4}$~s. Considering that an atmospheric shower takes about 0.3
seconds, and that the detector resolution is about 2.5~$\mu$s, it is
possible to determine the coincidence of these decays.

\begin{figure}
\includegraphics[width=\linewidth]{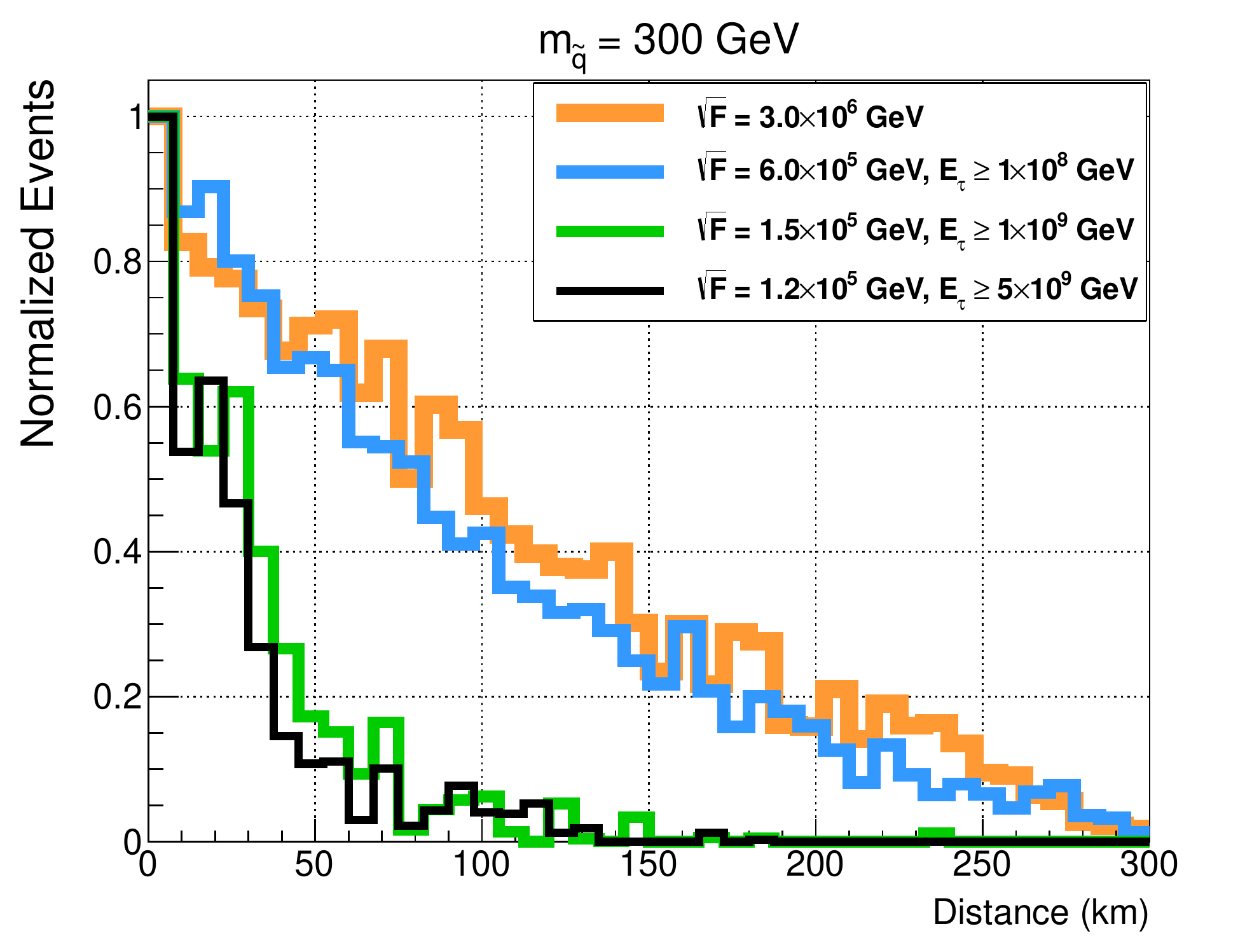}
\caption{Distance between two $\tau$ decays in the atmosphere, assuming
  same parameters as in the previous figure.}
\label{fig:dist} 
\end{figure}

Figure~\ref{fig:dist} shows the distance between the coincidence
showers, considering the same parameters as for the previous figure.
Given that the detector spacial resolution is about 0.75~Km, it is
clear that the majority of the coincident showers can be well distinguished. 

\section{\label{sec:conc} Discussion And Conclusions}

In order to understand the effect of the experimental energy
threshold, we showed our results for arbitrary lower energy cuts. As
shown in Table~\ref{tab:encut} , the lower threshold has a significant
effect on how well large fluorescence telescopes can probe the
supersymmetry breaking scale. 

While an experimental energy threshold
of $10^{18}$~eV allows JEM-EUSO to discover NLSP produced $\tau$s and
consequently probe a significant $\sqrt{F}$ region, a $10^{19}$~eV allows to set
limits in the production and $\sqrt{F}$
parameters. Table~\ref{tab:encut} shows that, for a maximum value of
$\sqrt{F}$, $\vartheta(10^5)$ of detectable events will go through
JEM-EUSO FOV. From these hundreds of events per year can be seen if the energy
threshold is set to $10^{17}$~eV, while tens for $10^{18}$ and more
than a year is needed to detect events for
$10^{19}$~eV. Although these numbers are for a maximized value of
$\sqrt{F}$, one can see from Figure~\ref{fig:encut} that there is a
significant range, in which $10^5 \lae \sqrt{F} \lae 10^6$~GeV,
that can be probed with the same order of events.

We have shown that large fluorescence telescopes such as JEM-EUSO have
the potential to indirectly detect NLSPs which are modeled in various
supersymmetry breaking scenarios. We show that  scenarios where the
supersymmetry breaking scale $\sqrt{F}$ is such that the NLSP will
decay close to or in the atmosphere can be probed by JEM-EUSO. 
Depending on the experimental energy threshold, scenarios with $10^5
\lae  \sqrt{F} \lae 10^6$ can be probed, and NLSPs can be indirectly
searched.

This work complements the direct probe for long-lived
NLSPs~\cite{abc,abcd}, where scenarios with $5 \times 10^6
\lae  \sqrt{F} \lae 5 \times 10^8$ can be directly probed by neutrino
telescopes. It is also complementary to searches for NLSP decays at the LHC.

\acknowledgments
IA was partially funded by 
the Brazilian National Counsel for Scientific Research (CNPq), and J.C.S was  
funded by the State of S\~{a}o Paulo Research Foundation (FAPESP).


\begin{thebibliography}{99}
\bibitem{abc} I.~Albuquerque, G.~Burdman and Z.~Chacko,
  Phys.\ Rev.\ Lett.\  {\bf 92}, 221802 (2004)
  [arXiv:hep-ph/0312197] .
\bibitem{abcd} I.~Albuquerque, G.~Burdman and Z.~Chacko,
  Phys.\ Rev.\  D {\bf 75}, 035006 (2007)
  [arXiv:hep-ph/0605120].
\bibitem{abckk} I.~F.~M.~Albuquerque, G.~Burdman, C.~A.~Krenke and B.~Nosratpour,
  Phys.\ Rev.\  D {\bf 78}, 015010 (2008)
  [arXiv:0803.3479 [hep-ph]].
\bibitem{gmsb} G.~F.~Giudice and R.~Rattazzi,
  Phys.\ Rept.\  {\bf 322}, 419 (1999)
  [hep-ph/9801271].
\bibitem{tese} Jairo Cavalcante de Souza, PhD. Thesis, 
  http://www.teses.usp.br/teses/disponiveis/,
  Universidade de S\~{a}o Paulo, in portuguese, S\~{a}o Paulo  (2012).
\bibitem{decice} I.~F.~M.~Albuquerque and J.Cavalcante de Souza.
arXiv:1210.5141 [hep-ph] (2012).
\bibitem{euso}   F.~Kajino {\it et al.}  [JEM-EUSO Collaboration],
  AIP Conf.\ Proc.\  {\bf 1367}, 197 (2011); http:\/\/jemeuso.riken.jp/en/index.html
\bibitem{wb} E.~Waxman and J.~N.~Bahcall,
Phys.\ Rev.\ D {\bf 59}, 023002 (1999); 
J.~N.~Bahcall and E.~Waxman,
Phys.\ Rev.\ D {\bf 64}, 023002 (2001).
\bibitem{pdg}  J. Beringer et al. (Particle Data Group), Phys. Rev. D86, 010001 (2012).
\bibitem{stbbn} 
  M.~Kawasaki, K.~Kohri, T.~Moroi and A.~Yotsuyanagi,
  Phys.\ Rev.\ D {\bf 78}, 065011 (2008)
  [arXiv:0804.3745 [hep-ph]].
\bibitem{earth} Adam Dziewonski, Earth Structure, Global, in The Encyclopedia of Solid Earth
Geophysics, edited by David E. James (Van Nostrand Reinhold, New York,
p. 331 (1989).
\bibitem{gandhi} R.~Gandhi, C.~Quigg, M.~H.~Reno and I.~Sarcevic,
Astropart.\ Phys.\  {\bf 5}, 81 (1996); 
R.~Gandhi, C.~Quigg, M.~H.~Reno and I.~Sarcevic,
Phys.\ Rev.\ D {\bf 58}, 093009 (1998). 
\bibitem{ina} M.~H.~Reno, I.~Sarcevic and S.~Su,
  Astropart.\ Phys.\  {\bf 24}, 107 (2005)
  [arXiv:hep-ph/0503030].
\bibitem{crotty} 
  P.~R.~Crotty,
  FERMILAB-THESIS-2002-54 (2002).
\bibitem{dutta} 
  S.~I.~Dutta, M.~H.~Reno and I.~Sarcevic,
  Phys.\ Rev.\ D {\bf 62}, 123001 (2000)
  [hep-ph/0005310].
\end{thebibliography}
\end{document}